\documentclass[aip,reprint]{revtex4-1}

\draft % marks overfull lines with a black rule on the right
\usepackage{bm}
\usepackage{amsmath}
\usepackage{graphicx}
\usepackage{multirow}
\usepackage{colortbl}
\usepackage{makecell}
\usepackage{tabularx}

\definecolor{tabcolor}{rgb}{.210,.10,.11}
\usepackage[unicode=true,pdfusetitle,
 bookmarks=true,bookmarksnumbered=false,bookmarksopen=false,
 breaklinks=false,pdfborder={0 0 0},pdfborderstyle={},backref=false,colorlinks=true]
 {hyperref}
\hypersetup{citecolor=blue}
\begin{document}

%\preprint{}

\title{Exceptional points in a topological photonic system} %Title of paper



\author{Junhua Dong}
\affiliation{Key Lab of advanced optoelectronic quantum architecture and measurement (MOE), School of Physics, Beijing Institute of Technology, Beijing 100081, China.}
\author{Qingmei Hu}
\affiliation{Key Lab of advanced optoelectronic quantum architecture and measurement (MOE), School of Physics, Beijing Institute of Technology, Beijing 100081, China.}
\author{Changyin Ji}
\affiliation{Key Lab of advanced optoelectronic quantum architecture and measurement (MOE), School of Physics, Beijing Institute of Technology, Beijing 100081, China.}
\author{Bingsuo Zou}
\affiliation{MOE $\&$ Guangxi Key Laboratory of Processing for Non-ferrous Metals and Featured Materials, School of Physical science and Technology, Guangxi University, Nanning 530004, China.}
\author{Yongyou Zhang}
\email[Author to whom correspondence should be addressed. Electronic mail: ]{yyzhang@bit.edu.cn}
%\email[Electronic mail: ]{yyzhang@bit.edu.cn}
\affiliation{Key Lab of advanced optoelectronic quantum architecture and measurement (MOE), School of Physics, Beijing Institute of Technology, Beijing 100081, China.}

%\thanks{}


\date{\today}

\begin{abstract}
Exceptional points as branch singularities describe peculiar degeneracies of non-Hermitian systems that do not obey energy conservation. This work shows that exceptional points can emerge in a topological photonic system, for example, the topological photonic waveguide coupled with two degenerate counter-propagation topological whispering gallery modes. Such a photonic architecture is designed by crystal-symmetry-protected topological photonic insulators based on air rods in conventional dielectric materials. The relevant exceptional point reveals the breaking of the parity-time symmetry, reflected by the change of the transmission-dip number in the optical transmission spectra of the system. Achieving exceptional points in topological photonic systems possibly opens a new avenue toward robust optical devices with exceptional-point-based unique properties and functionalities.
\end{abstract}

%\pacs{42.50.Pq 42.79.Gn 03.65.Vf}% insert suggested PACS numbers in braces on next line

\maketitle %\maketitle must follow title, authors, abstract and \pacs

\section{Introduction}
Hermiticity is required by plenty of physical models, if they are assumed to be energy conservative and time-reversal symmetric. However, non-Hermitian physics, as a counterpart of Hermitian physics \cite{Bender2002CQM,Jin2010physics}, has attracted a lot of interest in recent years, from quantum physics \cite{Bender1998Real, Moiseyev1998Resonances} to optics or photonics \cite{Jin2018incident, Zhen2020EP}. Non-Hermitian phenomena have been revealed to be able to dramatically alter the properties of a system. One of the best known examples is the so-called exceptional points (EPs), which are branch point singularities in the parameter space of the system. More than one eigenvalues at the EPs and associated eigenvectors coalesce simultaneously and therefore, the system becomes degenerate. If the system holds the parity-time ($\cal PT$) symmetry, the corresponding Hamiltonian can support purely real eigenvalue spectra \cite{Bender1998Real}. Moreover, the $\cal PT$-symmetric Hamiltonian can reach a spontaneously-broken regime through the phase transition, where eigenvalues become complex. Since the $\cal PT$-symmetric systems require gain and loss channels, the photonic structures are ideal platforms for exploring the non-Hermitian physics, for example, integrated photonic waveguides \cite{Guo2009WG}, coupled micro-resonators \cite{Yang2014Resonators}, optomechanical architectures \cite{Alu2017Optomechanical}, and so on \cite{Peschel2013Lattice, Khurgin2020EP}. With these architectures, researchers have achieved coherent laser absorber \cite{Douglas2011Laserabsorber}, unidirectional invisibility \cite{Alu2015Cloaking}, negative refraction \cite{Alu2014NegativeRfraction}, and anisotropic transmission resonances \cite{Alu2013Metamaterials} at the EPs. The singularity of EPs not only opens potentiality for advanced optical manipulations such as enhanced mode splitting \cite{weijian2017exceptional}, but also can be harnessed for mode discrimination in multimode laser cavity \cite{m2012large}. However, these photonic architectures have not yet been integrated with the burgeoning topological photonics field. It should be evidently interesting to study/achieve non-Hermitian physics within the topological photonics.
\begin{figure*}
  \centering
  \includegraphics[width=1\textwidth]{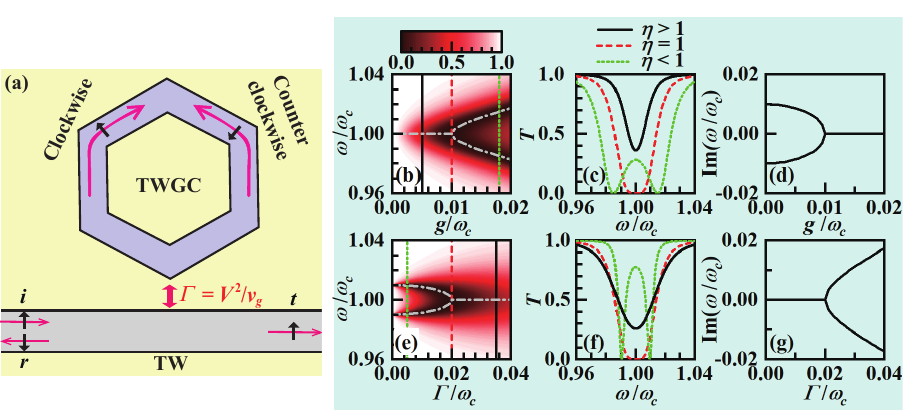}\\
  \caption{(a) Schematics of the TWGC side coupled with the bus TW. Both TWGC and bus TW support two oppositely-moving TESs with up and down spin, respectively. The coupling between the bus TW and TWGC is taken as $\delta$-type with the strength $V$. $V$ shows influence through the effective coupling $\varGamma\equiv V^2/v_g$ where $v_g$ is the TES group velocity in the bus TW. Variations of transmission spectra with the TWGM backscattering strength $g$ are in (b, c) and those with the effective coupling $\varGamma$ are in (e, f). The dash dot lines in (b, e) give the real parts of eigenvalues of the $\cal PT$ symmetric model in Eq.~\eqref{eq-PT}, denoting the position of the local transmission minimum. Variations of the imaginary parts of the eigenvalues in Eq.~\eqref{eq-PT} are in (d, g). The black solid, red dashed, and green dotted curves in (c, f) correspond to those in (b, e), respectively. In (b-d), $\varGamma=0.02\omega_c$ and in (e-g), $g=0.01\omega_c$.}
  \label{fig-model}
\end{figure*}

Topological photonics triggered by topological electronics \cite{hsieh2008a} can simulate a number of electrical topological phenomena, such as quantum (spin) Hall effect \cite{barik2018topological}, high-order topological insulators \cite{Xu2020Corner, Minkyung2020Corner},  Weyl semimetals \cite{yang2018ideal}, photonic topological valley Hall effect \cite{Yoshimi20}, and so on \cite{slobozhanyuk2017three, mittal2019photonic1}. The remarkable application of  topological photonic insulators (ToPIs) rests in the robustness of optical properties against perturbations. The topological optical interfaces are often used to design ideal waveguides for topological edge states (TESs) \cite{xia2017topological, ji2019transport} and the topological corners can work as the amazing optical cavities \cite{Xu2020Corner, Minkyung2020Corner}. Other types of robust optical devices have also been attempted, for example, topological lasers \cite{gal2018topological, Miguel2018topological} and perfect reflectors \cite{ji2020fragile}. Accordingly, the ToPIs have attracted extensive attention \cite{xia2017topological, wu2015scheme} and provide an extraordinary platform for exploring and understanding topological protection, as well as for the topological EPs.

Since EPs present a number of peculiar properties \cite{Douglas2011Laserabsorber, Alu2015Cloaking, Alu2014NegativeRfraction, Alu2013Metamaterials}, it is a rational expectation to introduce an EP into a topological photonic system, where the topology of the system can supply a protection for the EP against perturbations. Furthermore, one may expect that the whole system has no gain or loss, since introducing gain and loss into the system commonly leads to fabrication complexity.Accordingly, the present work focuses on the nontrivial EP in the topological photonic system, as a whole, without gain or loss. Here, the topological photonic waveguide coupled with two degenerate counter-propagation topological whispering gallery modes is taken as an example. Note that the designed structure is Hermitian as a whole, but the internal subsystem, i.e., the whispering-gallery cavity, can be regarded as non-Hermitian due to its energy exchange with the waveguide and the coupling between the two modes. In fact, trivial EPs widely exist in the systems without gain or loss, such as critical angle of the total internal reflection at the interface between two dielectric materials, cut-off frequency of a closed waveguide, and band edge of a photonic crystal \cite{Alu2019Review}. The emergent EP in the non-Hermitian subsystem reveals the breaking of the $\cal PT$ symmetry, reflected by the change of the transmission-dip number in the transmission spectra of the system in this work. Since the optical architecture is topological, the relevant EP should be robust against the system imperfections, such as waveguide bending, disorder, and rod missing.
%%\label{}
\section{Model and formulas}\label{model}
\begin{figure*}
  \centering
  \includegraphics[width=0.9\textwidth]{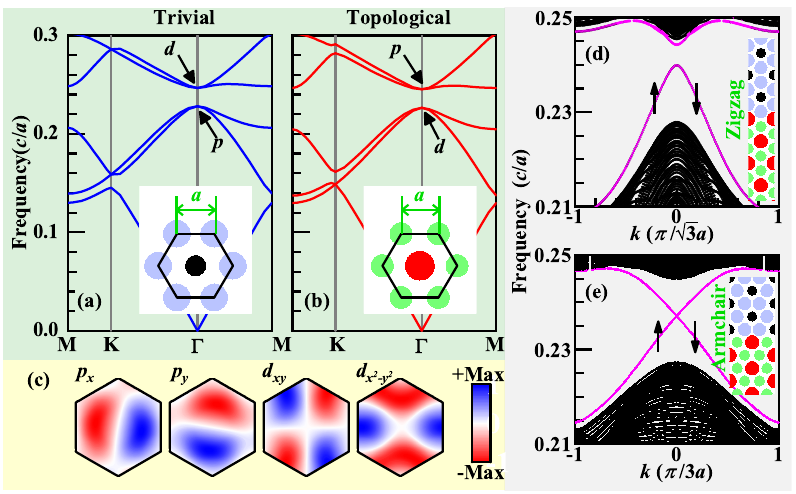}
  \caption{Band structures for (a) trivial photonic crystals (TrPIs) and (b) topological photonic crystals (ToPIs). The insets are their unit cells. (c) Field distributions of the $p_x$, $p_y$, $d_{xy}$, and $d_{x^2-y^2}$ orbitals at $\Gamma$ point. (d, e) Topological edge bands for the zigzag and armchair interfaces composed of the TrPI and ToPI in (a) and (b). The bulk (edge) states are denoted by black (magenta) dots. The rightward- and leftward-moving TESs carry the up and down pseudospin, respectively.}\label{fig-bands}
\end{figure*}
The architecture considered is composed of one topological waveguide (TW) and one topological whispering-gallery cavity (TWGC), see Fig.~\ref{fig-model}(a). They both are achieved by the topological interfaces of two topology-different photonic insulators. We define that the rightward-moving (leftward-moving) TES in the TW and the clockwise (counter clockwise) rotating TES in the TWGC carry the up (down) spin or pseudospin. Photons as carriers of information transporting in the bus waveguide can be effectively adjusted by the quantum emitters (coupled with the bus waveguide), for example, optical cavities, two-level atoms, and Jaynes-Cummings models \cite{pan2010experimental, tan2011entangling, dong2019transport}. The TWGC in Fig.~\ref{fig-model}(a) plays such a role. Owing to spin conservation, the rightward-moving (leftward-moving) TES in the bus TW only couples with the clockwise (counter clockwise) topological whispering-gallery mode (TWGM) in the TWGC with the strength $V$. This coupled architecture is described by the following Hamiltonian \cite{Zhang2014b, ji2019transport},
\begin{align}
H&=\sum_{\sigma=\pm}\left\{\int dx \hat \psi_\sigma^\dag (x)\hat \omega_\sigma(-i \partial_x ) \hat \psi_\sigma(x)+\omega_c \hat c_\sigma^\dag \hat c_\sigma \right\}\nonumber\\
&+ \sum_{\sigma=\pm} \int dx V\delta(x)\left[\hat c_\sigma^\dag\hat \psi_\sigma(x) + \hat \psi_\sigma^\dag(x)\hat c_\sigma\right] \nonumber\\
&+ g\left( \hat c_+^\dag \hat c_- + \hat c_-^\dag \hat c_+  \right),
\label{eq-H}
\end{align}
where $x$ is the coordinate along the bus waveguide and the TWGC is placed at the original point. For convenience, the Planck constant is set to be $\hbar=1$ henceforth. $\sigma=+\ (-)$ denotes the spin-up (down) TESs. $\hat c_\sigma$ $\left(\hat c_\sigma^\dag\right)$ is the annihilation (creation) operator of the TWGM with spin $\sigma$ and eigenfrequency $\omega_c$. Note that the clockwise and counter clockwise TWGMs are degenerate, between which the backscattering strength is measured by the parameter $g$. The value of $g$ can be controlled by designing the geometry of the TWGC. $\hat\psi_\sigma$ $\left(\hat\psi_\sigma^\dag\right)$ is the annihilation (creation) field operator for the TES with spin $\sigma$ in the bus TW, whose dispersion $\hat\omega_\sigma(-i\partial_x)$ can be linearized as $\omega_\sigma(k)=\omega_c+v_g(\sigma k- k_c)$, corresponding to $\omega_\sigma(\sigma k_c)=\omega_c$ when $k=\sigma k_c$. $v_g$ and $k$ are the group velocity and the wave vector of the TES, respectively. Here, the coupling between the bus TW and TWGC is taken as $\delta$-type with the strength $V$.

To derive the TES (photon) transmission, the following single-particle wave function for $H$ is adopted,
\begin{align}
|\Phi\rangle&=\sum_{\sigma=\pm}\int dx {\cal W}_\sigma(x)\hat \psi_\sigma^\dag (x)|\emptyset\rangle\nonumber\\
&+\sum_{\sigma=\pm}{\cal C}_\sigma\hat c_\sigma^\dag |\emptyset\rangle
\label{eq_Phi}
\end{align}
where $|\emptyset\rangle$ represents the vacuum state with zero photon in the TWGC or bus TW. ${\cal W}_\sigma(x)$ and ${\cal C}_\sigma$ are the wave function of the spin-$\sigma$ TES in the bus TW and the excitation amplitude of the spin-$\sigma$ TWGM. Since the TWGC is placed at the original point, ${\cal W}_\sigma(x)$ can be constructed as the following form through the system reflection and transmission coefficients, $r$ and $t$, i.e.,
\begin{align}
{\cal W}_+(x)=e^{ikx}[\theta(-x)+t\theta(x)], \quad
{\cal W}_-(x)=re^{-ikx}\theta(-x),
\label{eq-W}
\end{align}
where $\theta(x)$ is the unit step function. Substituting Eqs.~\eqref{eq-H} and \eqref{eq_Phi} into the Schr\"odinger equation,
\begin{align}
i \frac{\partial}{\partial t}|\Phi\rangle=H|\Phi\rangle,
\end{align}
leads to the coupled equation set for ${\cal W}_\sigma(x)$ and ${\cal C}_\sigma$ as follows:

\begin{align}
&i\frac{\partial}{\partial t}{\cal W}_+(x)=\hat \omega_+(-i \partial_x ) {\cal W}_+(x) + V\delta(x){\cal C}_+, \nonumber\\
&i\frac{\partial}{\partial t}{\cal W}_-(x)=\hat \omega_-(-i \partial_x ) {\cal W}_-(x) + V\delta(x){\cal C}_-,\nonumber\\
&i\frac{\partial}{\partial t}{\cal C}_+=\omega_c{\cal C}_+ +g{\cal C}_- + V{\cal W}_+(0),\nonumber\\
&i\frac{\partial}{\partial t}{\cal C}_-=\omega_c{\cal C}_- +g{\cal C}_+ + V{\cal W}_-(0).
\end{align}
\label{eq-Dy}

In the steady state case, ${\cal W}_\sigma(x)$ and ${\cal C}_\sigma$ satisfy the relation,
\begin{align}
i\frac{\partial}{\partial t}{\cal W}_\sigma(x) = \omega {\cal W}_\sigma(x), \qquad i\frac{\partial}{\partial t}{\cal C}_\sigma = \omega {\cal C}_\sigma,
\end{align}
with the oscillation frequency $\omega$. Further substituting $\omega_\sigma(-i\partial_x)=\omega_c - v_g k_c -i\sigma v_g\partial_x$ and the wave functions Eq.~\eqref{eq-W} into Eq.~\eqref{eq-Dy}, one can find the transmission coefficient $t$, i.e.,
\begin{align}\label{eq-t}
t=\frac{\left[\omega-\left(\omega_c-i\frac{\varGamma}{2}\right)\right]\left[\omega-\left(\omega_c+i\frac{\varGamma}{2}\right)\right]-g^2}{\left(\omega-\omega_c+i\frac{\varGamma}{2}\right)^2-g^2},
\end{align}
where $\varGamma\equiv V^2/v_g$ describes the effective coupling between the bus TW and TWGC, see Fig.~\ref{fig-model}(a). The corresponding transmissivity reads
\begin{align}
T=|t|^2.
\end{align}
The transmission spectra are plotted as functions of $g$ and $\varGamma$ in Figs.~\ref{fig-model}(b-c) and (e-f). One or two minimum transmission points (MTPs) can be observed in the frequency domain, determined by the ratio $\eta\equiv \varGamma/2g$. For $\eta<1$ there are two MTPs at which the transmissivities are exactly equal to zero, while for $\eta>1$ there is only one at which the transmissivity is greater than zero, see Figs.~\ref{fig-model}(c, f). Two regions are separated by the transition point of $\eta=1$, as demonstrated by the grey dash-dotted curves in Figs.~\ref{fig-model}(b, e). Such a transition point, intuitively, is just the EP in the present topological architecture. It can be confirmed by mapping the numerator of Eq.~\eqref{eq-t} to the following two-level $\cal PT$ symmetric Hamiltonian,
\begin{align}
H_{\cal {PT}}=\left(\begin{array}{cc}
\omega_c - i\frac{\varGamma}{2} & g \\
g & \omega_c + i\frac{\varGamma}{2}
\end{array}\right),
\label{eq-PT}
\end{align}
where $\varGamma$ measures the gain and loss rates of the two levels. Though the whole coupled TW-TWGC structure is Hermitian [see Eq.~\eqref{eq-H}], the TWGC as a subsystem could be non-Hermitian since it couples with the bus TW. A simple derivation for this effective Hamiltonian is provided in Appendix. The zero-transmission points in Eq.~\eqref{eq-t} are exactly determined by the eigenfrequencies of $H_{\cal {PT}}$, reading as
\begin{align}\label{eq_w}
\omega_\pm=\omega_c\pm\sqrt{g^2-\frac{\varGamma^2}{4}}.
\end{align}
The evaluation of the imaginary parts of $\omega_\pm$ are given in Figs.~\ref{fig-model}(d, g). When $\eta$ increases from 0 to 1, the two eigenfrequencies coalesce and the EP is achieved at $\eta=1$. In the region with $\eta>1$, the $\cal PT$ symmetry is spontaneously broken \cite{Alu2019Review, Ganainy2019Review}, resulting in complex eigenfrequencies, see Figs.~\ref{fig-model}(d, g). The interesting connection between the Hamiltonians in Eqs.~\eqref{eq-H} and \eqref{eq-PT} could be argued as follows. The clockwise TWGM in the TWGC gains excitation or energy from the incident rightward-moving TES, then transfers energy to the counter clockwise TWGM through backscattering, and finally the counter clockwise TWGM losses energy by coupling with the leftward-moving TES. Here, the rates for gain and loss are exactly equal to each other, namely, $\varGamma$, which is attributed to the spin conservation in the coupling between the bus TW and TWGC and the time-reversal symmetry of the whole system. The two modes in Eq.~\eqref{eq-PT} just correspond to the clockwise and counter clockwise TWGMs. Their coupling with the TESs in the bus TW is responsible for the emergence of the $\cal PT$ symmetric Hamiltonian in Eq.~\eqref{eq-PT}, which confirms that the transition point at $\eta=1$ is the EP in the present system. The EP is reflected by the coalescence of the two MTPs as $\eta$ increases from 0 to that greater than 1, see Figs.~\ref{fig-model}(b,c) and (e,f). Furthermore, the topology of the system would provide protection for the EP against perturbations. The achievement of the EP on a Hermitian topological platform could bring about convenience for optical applications. Next section will show a practical example for achieving the topological EP, based on ToPIs.
\section{Achieve topological EPs}\label{simulation}
\begin{figure*}
  \centering
  \includegraphics[width=1\textwidth]{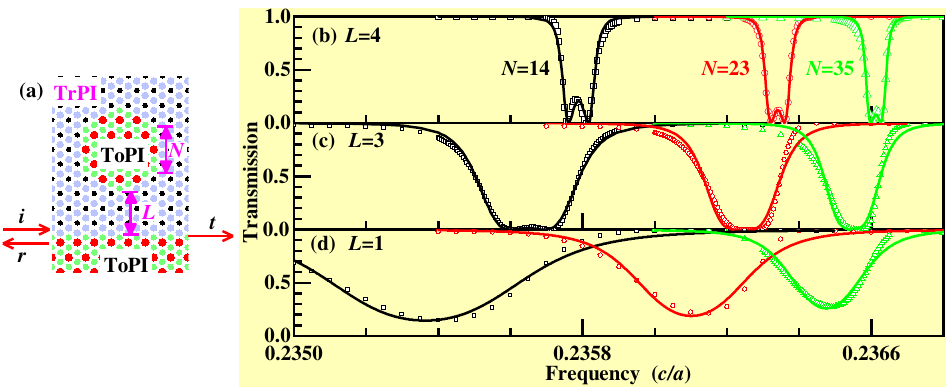}
  \caption{Schematic of the TW-TWGC architecture. The radii of air rods are identical to those given in Fig.~\ref{fig-bands}. (b-d) Transmission spectra of the TW-TWGC architecture for different $N$ and $L$. Hollow scatter dots are calculated by the finite element method within Comsol code and the solid lines present the theoretical fitting using Eq.~\eqref{eq-t} with the fitted parameters listed in Table \ref{table-fit}.}\label{fig-EP}
\end{figure*}

The ToPIs used here are the crystal-symmetry-protected systems \cite{simon2017crystalline, maxim2018far}, based on the hexagonal air-rod lattice in the silicon plate with relative dielectric constant $\varepsilon_r=11.7$, whose unit cells are denoted in Figs.~\ref{fig-bands}(a, b). The silicon-photonic crystals are widely fabricated in experiments with advanced micro/nano-processing technology \cite{a2001silicon}. If the central air rod is smaller than those around it, the lattice behaves as a trivial photonic insulator (TrPI), otherwise the lattice behaves as a topological one \cite{wu2015scheme, ji2019transport}, reflected by the band inversion of $p$ and $d$ orbitals at $\Gamma$ point, see Figs.~\ref{fig-bands}(a, b). The $C_{6v}$ symmetry of the system leads to the degeneracy of the two $p$ or two $d$ bands at $\Gamma$ point, whose field distributions are plotted in Fig.~\ref{fig-bands}(c). The $p_x$ and $p_y$ ($d_{xy}$ and $d_{x^2-y^2}$) orbitals are the bases of the two-dimensional irreducible representation $E_1$ ($E_2$) of $C_{6v}$. With them the pseudospin up and down states could be constructed as \cite{wu2015scheme, ji2019transport}
\begin{align}\label{eq_spin}
p_\pm = (p_x\pm ip_y)/\sqrt{2},\quad d_\pm=(d_{xy}\pm id_{x^2-y^2})/\sqrt{2}.
\end{align}
The corresponding time reversal operator is expressed as ${\cal T}=-\sigma_y{\cal K}$ with complex conjugate operator $\cal K$ and Pauli matrix $\sigma_y$ operating on $p_\pm$ and $d_\pm$ \cite{wu2015scheme, ji2019transport}. On the bases of $p_\pm$ and $d_\pm$, ${\cal T}^2=-1$ is responsible for the nontrivial topology in Fig.~\ref{fig-bands}(b). In order to match the band gaps the radii of the air rods for the TrPI and ToPI are set to be $0.32a$ for the black, $0.42a$ for the light blue, $0.45a$ for the red, and $0.35a$ for the green, where $a$ is the distance between the adjacent rods, see the insets in Figs.~\ref{fig-bands}(a, b). The band gap is $(0.2281, 0.2468)\frac{c}{a}$ for the TrPI and is $(0.2265, 0.2453)\frac{c}{a}$ for the ToPI ($c$ is the speed of light in vacuum), matching well. The bands of the TESs for the zigzag and armchair interfaces between the TrPI and ToPI are shown in Figs.~\ref{fig-bands}(d, e), where the rightward-moving (leftward-moving) TESs are assumed to carry the up (down) pseudospin. Owing to the breaking of the $C_{6v}$ symmetry on the interfaces, there exists a gap at the cross point of the pseudospin-up and -down dispersions, which indicates that the crystal-symmetry could not provide perfect protection for the TES against perturbations. The value of this gap can be adjusted by controlling the geometry of the interface \cite{wu2015scheme}. On the other hand, only when the frequencies of the perturbation-induced states fall into the bulk band gap, the perturbations would exert a strong influence on the transport of the TESs \cite{ji2019transport}. Consequently, these two aspects guarantee that the TESs constructed by the crystal-symmetry-protected topological insulators are immune to most of common perturbations, such as waveguide bending, rod missing, and local disorder \cite{ji2020fragile}. In fact, this characteristic will be used to adjust the backscattering strength, $g$, between the clockwise and counter clockwise TWGMs.  Figures \ref{fig-bands}(d, e) show that this gap value is much smaller for the armchair interface than that for the zigzag one in the present crystal-symmetry-protected topological systems.
\begin{table}
 \centering
 \caption{Fitted parameters for the transmission spectra in Figs.~\ref{fig-EP}(b-d) and corresponding $\eta=\varGamma/2g$. The $\cal PT$ transition is at $\eta=1$.}\label{table-fit}
  \renewcommand\arraystretch{1.2}
  \begin{tabular}
  {p{0.6cm}<{\centering}|p{0.8cm}<{\centering}|p{1.5cm}<{\centering}|p{1.8cm}<{\centering}|p{1.8cm}<{\centering}!{\vrule width1.2pt} p{1.0cm}<{\centering}}
  %{c|c|c|c|c|c}
  \Xhline{1.5pt}
   $L$ &  $N$& $\omega_c / (c/a)$ & $\varGamma / (10^{-4}c/a)$ & $g / (10^{-4}c/a)$ & $\eta$\\
   \Xhline{1.5pt}
   \multirow{3}{*}{4} & 14 & 0.235789  & 0.418410  & 0.342990 & 0.61\\
   \cline{2-6}
                            & 23 & 0.236341  & 0.364198 & 0.262572 & 0.69\\
   \cline{2-6}
                            & 35 & 0.236613  & 0.311221 & 0.209549 & 0.74\\
   \Xhline{1.5pt}
   \multirow{3}{*}{3} & 14 & 0.235648  & 1.818314  & 1.02660 & 0.89\\
   \cline{2-6}
                            & 23 & 0.236242  & 1.505573 & 0.767392 & 0.98\\
   \cline{2-6}
                            & 35 & 0.236547  & 1.186634 & 0.557715 & 1.06\\
   \Xhline{1.5pt}
   \multirow{3}{*}{1} & 14 & 0.235364 & 5.864978  & 1.960157 & 1.50 \\
   \cline{2-6}
                            & 23 & 0.236101 & 3.582460  & 1.122943 &  1.60\\
   \cline{2-6}
                            & 35 & 0.236475 & 2.539101  & 0.723024 & 1.76\\
    \Xhline{1.5pt}
\end{tabular}
\end{table}
\begin{figure*}
  \centering
  \includegraphics[width=0.95\textwidth]{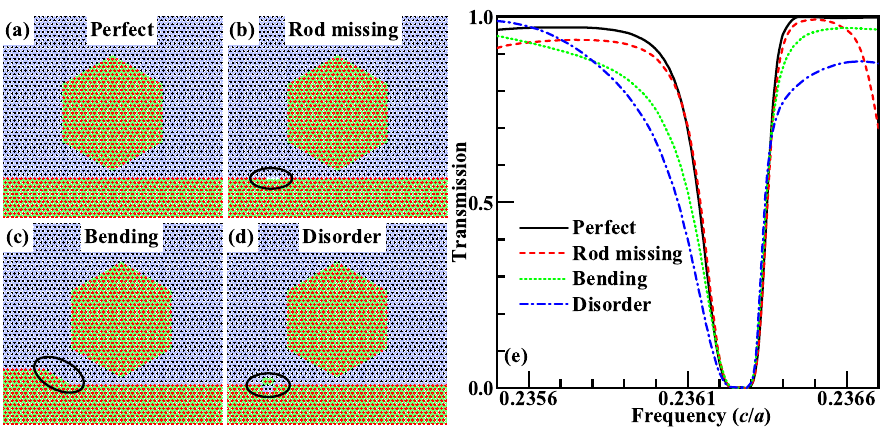}
  \caption{Schematic diagrams of TW-TWGC architectures with (a) perfect interface, (b) rod missing, (c) bending interface, and (d) local disorder. The perturbations are marked by the ellipses. (e) Transmission spectra for the four TW-TWGC architectures given in (a-d). The radii of air rods are identical to those given in Fig.~\ref{fig-bands} but here $N=23$ and $L=3$ are adopted.}\label{fig4-disorder}
\end{figure*}
The zigzag interface is designed as the bus TW, while the closed armchair one is used to achieve the TWGC, referred to Fig.~\ref{fig-EP}(a). The air-rod number along one side of the TWGC is denoted as $N$ and that between the TW and TWGC is $L$. Owing to the pseudospin conservation, the incident pseudospin-up (reflected pseudospin-down) wave only couples with the clockwise (counter clockwise) TWGM, see Figs.~\ref{fig-model}(a) and \ref{fig-EP}(a). The coupling strength $\varGamma$ between the bus TW and TWGC and the backscattering strength $g$ between the two TWGMs both decrease with the increasing $L$. The former is intuitive, since $L$ measures the distance between the bus TW and TWGC. The later is attributed to that the bus TW shows weaker and weaker perturbation on the armchair interface in the TWGC as $L$ increases. On the other hand, as $N$ increases they both decrease also. The former originates from the extension of the TWGMs and the later is attributed to the overlap decrease of the TWGMs across the TWGC. These variations are not linear and therefore, $\eta$ is different for different $L$ and $N$, which can be confirmed by fitting the transmission spectra in Figs.~\ref{fig-EP}(b-d) using Eq.~\eqref{eq-t}. In addition, the eigenfrequency $\omega_c$ exhibits a blueshift for increasing $L$ and $N$, which is due to the decrease of the bus TW influence and that of the TWGM overlapping across the TWGC. In Figs.~\ref{fig-EP}(b-d), three different $L$ and three different $N$ are adopted, for which the hollow scatters are calculated from the finite element method (FEM) within Comsol code and the solid curves are their fitted ones by Eq.~\eqref{eq-t} with the parameters listed in Table~\ref{table-fit}.

For $L=4$ the spectra present two MTPs where the transmission is zero. Since $g$ decreases when $N$ increases, the two MTPs in the spectra become closer but do not coalesce, see Fig.~\ref{fig-EP}(b), which tells that the $\cal PT$ symmetry is satisfied, reflected by $\eta<1$, see the rows with $L=4$ in Table.~\ref{table-fit}. Obviously, the theoretical fitting is in good agreement with the numerical data from the FEM. As $L$ decreases to $L=3$,  both $\varGamma$ and $g$ increase, leading to the changes of the spectral line shapes, as well as $\eta$, see Fig.~\ref{fig-EP}(c) and the rows with $L=3$ in Table.~\ref{table-fit}. For the spectrum with $L = 3$ and $N = 23$ and that with $L = 3$ and $N = 35$, $\eta=0.98$ and 1.06, respectively, both of which approach the EP point where $\eta=1$, confirmed by the zero-transmission at the MTP in the spectra, see the red and green curves in Fig.~\ref{fig-EP}(c).

Decreasing $L$ to $L=1$, both $\varGamma$ and $g$ increase further, resulting in that the three spectra for $N=14$, 23, and 35 all hold only one MTP, see Fig.~\ref{fig-EP}(d). The non-zero transmission at the MTP tells that the $\cal PT$ symmetry is broken, corresponding to $\eta>1$. This is confirmed by the fitted parameters in the rows with $L=1$ in Table.~\ref{table-fit}. The weak mismatch between the FEM data and theoretical fitting in Fig.~\ref{fig-EP}(d) should be attributed to that the $\delta$-type interaction is not well enough to cover the coupling between the bus TW and TWGC when $L$ is very small. Instead, an extended one can be used, for example, Gaussian function \cite{Zhang2014b, ji2019transport}. Considering complex coupling functions would bring about difficulties for analyzing the physics, the $\delta$-type coupling is always used here, whose availability is confirmed by the acceptable fitting results in Figs.~\ref{fig-EP}(b-d). If fixing $N=23$, one can definitely observe the $\cal PT$ phase transition, reflected by the number of the MTPs varies from two to one as $L$ decreases from $L=4$ to $L=1$. As a result, the EP can be achieved in the present topological photonic system.

Since the system is topological, such an EP should be immune to most of perturbations, for example, waveguide bending, rod missing, and local disorder, as long as the eigenfrequencies of their induced optical modes is out of the topological band gap \cite{ji2020fragile}. To show this, we take the case with $N=23$ and $L=3$ as an example. The architectures of the rod missing, waveguide bending, and local disorder are shown in Figs.~\ref{fig4-disorder}(b-d), respectively, and the case without perturbation (i.e., the perfect case) is provided in Fig.~\ref{fig4-disorder}(a) for comparison. The transmission spectra of these four cases are summarized in Fig.~\ref{fig4-disorder}(e) where the black solid curve is identical to the red circle dot curve in Fig.~\ref{fig-EP}(c) with $\eta=0.98$. Obviously, the transmission spectra maintain the line shape near the transmission valley when the perturbations are introduced, comparing the red dashed, green dotted, and blue dash-dotted curves with the black solid one. Accordingly, the EP in this system is topologically protected and robust against the perturbations.

\section{Conclusion}\label{conclusion}

To summarize, a nontrivial exceptional point is achieved in a topological photonic system, that is, the hexagonal optical architecture comprised of a topological photonic waveguide and a topological whispering gallery cavity. The zigzag topological interface is taken as the topological bus waveguide and the closed armchair one is regarded as the topological whispering-gallery cavity. The effective coupling strength between the waveguide and cavity, $\varGamma$, plays the roles of gain and loss rates in the subsystem of cavity, which guarantees the realization of exceptional points. $\varGamma$ and the backscattering strength between the two topological whispering-gallery modes (clockwise and counter clockwise), $g$, together determine the TES transmission. The parameter $\eta=\varGamma/2g=1$ gives the exceptional point. When $\eta<1$, the parity-time symmetry of the system is satisfied, otherwise is broken. $\varGamma$ and $g$ can be adjusted by controlling the distance between the waveguide and cavity or the cavity size. The achievement of exceptional points in topological photonic systems paves a way for robust optical devices with exceptional-point-based unique properties and functionalities.
\begin{acknowledgments}
 The authors would like to thank the support from National Natural Science Foundation of China (Grant No. 12074037)
\end{acknowledgments}

\appendix
\setcounter{equation}{0}
\renewcommand\theequation{A\arabic{equation}}
\section*{Appendix: Derivation of $H_{\cal {PT}}$ in Eq.(9)}

For convenience, we transform the Hamiltonian in Eq.~\eqref{eq-H} into the momentum space, i.e.,
\begin{align}
H=H_c+\int {dk\over2\pi} \sum_{\sigma=\pm}\left[\omega_{k\sigma}
\hat \psi_{k\sigma}^\dag\hat \psi_{k\sigma}+ V\left(\hat c_\sigma^\dag\hat \psi_{k\sigma} + \hat \psi_{k\sigma}^\dag\hat c_\sigma\right)\right],
\label{eq-Hk}
\end{align}
where $H_c=\sum_{\sigma} \omega_c \hat c_\sigma^\dag \hat c_\sigma+g\left( \hat c_+^\dag \hat c_- + \hat c_-^\dag \hat c_+  \right)$ and $\hat \psi_{k\sigma}=\int dx\hat \psi_\sigma(x)e^{-ikx}$. Using Eq.~\eqref{eq-Hk}, one can find the motion equations for $\hat \psi_{k\sigma}$ and $\hat c_\sigma$ as follows,
\begin{align}
\dot{\hat \psi}_{k\sigma}&=-i\omega_{k\sigma}\hat \psi_{k\sigma}-iV\hat c_\sigma,\label{eq-psisigma}\\
\dot{\hat c}_\sigma&=-i[\hat c_\sigma,\ H_c]-iV\int {dk\over2\pi}\hat \psi_{k\sigma}.\label{eq-csigma}
\end{align}
If the incident field takes $\sigma=+$, using Eq.~\eqref{eq-psisigma} it is intuitive to express $\hat \psi_{k+}$ at time $t$ by its initial field $\hat \psi_{k\sigma}(0)$ at time $t_0$, i.e.,
\begin{align}
\hat \psi_{k+}(t)&=\hat \psi_{k+}(0)e^{-i\omega_{k+}(t-t_0)} - iV\int_{t_0}^t dt' \hat c_+(t')e^{-i\omega_{k+}(t-t')},
\label{eq-psiku}
\end{align}
while $\hat \psi_{k-}$ at time $t$ can be expressed by its final field $\hat \psi_{k-}(t_1)$ at time $t_1$,
\begin{align}
\hat \psi_{k-}(t)&=\hat \psi_{k-}(t_1)e^{-i\omega_{k-}(t-t_1)} + iV\int_t^{t_1} dt' \hat c_-(t')e^{-i\omega_{k-}(t-t')}.
\label{eq-psikd}
\end{align}
Substituting Eqs.~\eqref{eq-psiku} and \eqref{eq-psikd} into Eq.~\eqref{eq-csigma}, one can obtain
\begin{align}
\dot{\hat c}_+&=-i[\hat c_+,\ H_c]-{\varGamma\over2}\hat c_+ - iV\int {dk\over2\pi}\hat \psi_{k+}(0)e^{-i\omega_{k+}(t-t_0)},\label{eq-cu}\\
\dot{\hat c}_-&=-i[\hat c_-,\ H_c]+{\varGamma\over2}\hat c_- - iV\int {dk\over2\pi}\hat \psi_{k-}(t_1)e^{-i\omega_{k-}(t-t_1)}.\label{eq-cd}
\end{align}
after some algebraic derivations. Here, the following relations
\begin{align}
&\int \!\! dk e^{-i\omega_{k\sigma}(t-t')}={2\pi\over v_g}\delta(t-t'),\nonumber\\
&\int_{t_0}^t \!\! dt' f(t')\delta(t-t')=\int_t^{t_1} \!\! dt' f(t')\delta(t-t')={1\over2}f(t).
\end{align}
are used. If we cast the $\varGamma$ terms in Eqs.~\eqref{eq-cu} and \eqref{eq-cd} into the Hamiltonian $H_c$ and neglect the waveguide fields, one can immediately get the effective Hamiltonian $H_{\cal {PT}}$ in Eq.~\eqref{eq-PT}.

%%\begin{widetext}
%%$$\mbox{put long equation here}$$
%%\end{widetext}
%
%
%
%
%% If you have acknowledgments, this puts in the proper section head.
%%\begin{acknowledgments}
%% Put your acknowledgments here.
%%\end{acknowledgments}
%
%% Create the reference section using BibTeX:

\section*{data availability}
The data that support the findings of this study are available from the corresponding author upon reasonable request.
\section*{references}
\bibliography{refs}

\end{document}